\documentclass[11pt]{revtex4}
\usepackage{amssymb}
\usepackage{latexsym}
\usepackage{epsfig}                                   %27 Jan 2013%
\usepackage{amsmath}

\begin{document}
\title{Emergence of spacetime dynamics in entropy corrected and braneworld models}
\author{A. Sheykhi $^{1,2}$\footnote{asheykhi@shirazu.ac.ir}, M. H. Dehghani$^{1,2}$ 
\footnote{mhd@shirazu.ac.ir} and S. E. Hosseini $^{1}$}
\affiliation{$^1$  Physics Department and Biruni Observatory, College of
Sciences, Shiraz University, Shiraz 71454, Iran\\
         $^2$  Center for Excellence in Astronomy and Astrophysics (CEAA-RIAAM) Maragha, P. O. Box 55134-441, Iran}

\begin{abstract}
A very interesting new proposal on the origin of the cosmic expansion was
recently suggested by Padmanabhan [arXiv:1206.4916]. He argued that the
difference between the surface degrees of freedom and the bulk degrees of
freedom in a region of space drives the accelerated expansion of the
universe, as well as the standard Friedmann equation through relation $%
\triangle V=\triangle t(N_{\mathrm{sur}}-N_{\mathrm{bulk}})$. In
this paper, we first present the general expression for the number
of degrees of freedom on the holographic surface,
$N_{\mathrm{sur}}$, using the general entropy corrected formula
$S=\frac{A}{4 L_{p}^2}+s(A)$. Then, as two example, by applying
the Padmanabhan's idea we extract the corresponding Friedmann
equations in the presence of power-law and logarithmic correction
terms in the entropy. We also extend the study to RS II and DGP
branworld models and derive successfully the correct form of the
Friedmann equations in these theories. Our study further supports
the viability of Padmanabhan's proposal.
\end{abstract}

\maketitle

\section{Introduction\label{Intr}}

Among all the fundamental forces of nature, clearly gravity is the most
universal. However, the nature and the origin of gravity has not known very
well yet. According to the equivalence principle, gravity is just the
dynamics of spacetime. This implies that gravity is an emergent phenomenon.
Indeed, the idea that gravity and spacetime geometry are emergent is widely
accepted nowadays. The universality of gravity also indicates that its
emergence should be understood from general principles that are independent
of the specific details of the underlying microscopic theory.

An interesting new idea on the origin of gravity was proposed by Verlinde
\cite{Ver} who claimed that gravity may be not a fundamental interaction but
should be interpreted as an entropic force caused by changes of entropy
associated with the information on the holographic screen. Applying the
holographic principle and the equipartition law of energy, Verlinde obtained
the Newton's law of gravitation. A relativistic generalization of this
argument directly leads to the Einstein equations. Similar discoveries were
also made by Padmanabhan \cite{Pad0}. He observed that the equipartition law
of energy for the horizon degrees of freedom combining with the
thermodynamic relation $S=E/2T$, also leads to the Newton's law of gravity,
where $S$ and $T$ are entropy and temperature associated with the horizon,
respectively, and $E$ is the active gravitational mass producing the
gravitational acceleration in the spacetime. This may imply that the entropy
is to link general relativity with the statistical description of unknown
spacetime microscopic structure when the horizon is present. It therefore
provides evidence for the fact that gravity can emerge from a microscopic
description that doesn't know about its existence.

The deep connection between thermodynamics and gravity has now well
established through a numerous and complementary theoretical investigations
\cite{Padm}. In \cite{Jac} Jacobson derived the Einstein field equations of
general relativity by applying the Clausius relation $\delta Q=T\delta S$ on
the horizon of spacetime, here $\delta S$ is the change in the entropy and $%
\delta Q$ and $T$ are the energy flux across the horizon and the Unruh
temperature seen by an accelerating observer just inside the horizon. Also,
by applying the Clauius relation to the apparent horizon of the
Friedmann-Robertson-Walker (FRW) universe, the corresponding Friedmann
equations can be derived in Einstein, Gauss-Bonnet and more general Lovelock
gravity \cite{CaiKim}. These kinds of studies were also generalized to the
brane cosmology, where it was shown that the differential form of the
Friedmann equation on the brane can be transformed to the first law of
thermodynamics on the apparent horizon \cite{SheyW1,SheyW2,CaiCao}. Indeed,
studies on the connection between gravity and thermodynamics have got a lot
of attentions in recent years (see \cite
{Elin,Akbar1,Akbar2,Par,Fro,Dan,Kot,Padm2,Cai2,Sheykhi1,Sheykhi2} and
references therein).

It is important to note that in most of these investigations, one
considers the gravitational field equations as the equations of
emergent phenomenon, but leave the spacetime as a pre-existed
background geometric manifold. Obviously, it is difficult to
regard the spacetime itself as an emergent structure. Whereas,
there are some conceptual difficulties associated with this idea.
For example, it is very hard to think that the time used to
describe the evolution of dynamical variables is emergent from
some pre-geometric variables and the space around finite
gravitational systems is emergent. Very recently, Padmanabhan
\cite{Pad1} suggested a new proposal for resolving these
difficulties by considering the emergence of spacetime in
cosmology. This is due to the fact that the cosmic time of a
geodesic observer provides a special choice of time variable, to
which the observed cosmic microwave background radiation is
homogeneous and isotropic, and the spatial expansion of our
universe can be regarded as the consequence of emergence of space.
According to Padmanabhan's idea \textit{the cosmic space is
emergent as the cosmic time progresses}. In addition, he argued
that the expansion of the universe is due to the difference
between the surface degrees of freedom and the bulk degrees of
freedom in a region of emerged space and successfully derived the
dynamical equation of a FRW universe. This approach provides a
novel paradigm to study the emergence of space and cosmology, and
has far reaching implications. Following \cite{Pad1}, Cai obtained
the Friedmann equation of a higher dimensional FRW universe. By
properly modifying the effective volume and the number of degrees
of freedom on the holographic surface from the entropy formulas of
static spherically symmetric black holes, he also derived
successfully the corresponding dynamical equations of the universe
in Gauss-Bonnet and more general Lovelock cosmology \cite{Cai1}.
In a similar way, the authors of \cite{Yang} derived the Friedmann
equations of a flat FRW universe in Gauss-Bonnet and Lovelock
cosmology from the generalized law governing the emergence of the
space \cite{Yang}. Instead of modifying the number of degrees of
freedom on the holographic surface of the Hubble sphere and the
volume increase, in \cite{Yang}, it is assumed that the volume
increase is
proportional to a function $f(\triangle N)$. Here $\triangle N=N_{\mathrm{sur%
}}-N_{\mathrm{bulk}}$, where $N_{\mathrm{sur}}$ is the number of degrees of
freedom on the boundary and $N_{\mathrm{bulk}}$ is the number of degrees of
freedom in the bulk. When the volume of the spacetime is constant, the
function $f(\triangle N)$ is equal to zero.

In general the entropy associated with the holographic surface is a function
of its area. Since the concept of number of degrees of freedom on the
holographic surface is closely related to the entropy, any modification of
the entropy expression leads to a particular number of degrees of freedom on
the holographic surface. In this paper, we would like to consider the
general form of the entropy expression in the presence of quantum correction
terms and derive the corresponding Friedmann equations by determining $N_{%
\mathrm{sur}}-N_{\mathrm{bulk}}$. We also apply the idea proposed in \cite
{Pad1} to the braneworld scenarios. By calculating the number of degrees of
freedom on the Hubble horizon of the brane we extract successfully the
Friedmann equations in Randall-Sundrum II (RS II) and Dvali, Gabadadze,
Porrati (DGP) braneworld models. Our study may indicate that the novel
proposal of Padmanabhan \cite{Pad1} is powerful enough to apply for deriving
the dynamical equations in other gravity theories.

The outline of our paper is as follows. In the next section, we present the
general formalism for deriving the dynamical equation of spacetime by
determining the difference between the bulk and the boundary degrees of
freedom. In section III, we derive the power-law and logaritmic
entropy-corrected Friedmann equations of FRW universe by using Padmanabhan
new idea. Then, in section IV, we extend the study to RS II and DGP
braneworld models and derive successfully the corresponding Friedmann
equations. The last section is devoted to some closing remarks.
%%%%%%%%%%%%%%%%%%%%%%%%%%%%%%%%%%%%%%%%%%%%%%%%%%%%%

\section{The General formalism\label{Gen}}

According to Padmanabhan's observation, the number of degrees of freedom on
the spherical surface of Hubble radius $H^{-1}$ is given by \cite{Pad1}
\begin{equation}
N_{\mathrm{sur}}=4S=\frac{A}{L_{p}^{2}}=\frac{4\pi }{L_{p}^{2}H^{2}},
\label{Nsur}
\end{equation}
where $L_{p}$ is the Planck length, $A=4\pi H^{-2}$ represents the area of
the Hubble horizon and $S$ is the entropy which obeys the area law. The bulk
degrees of freedom obey the equipartition law of energy,
\begin{equation}
N_{\mathrm{bulk}}=\frac{2|E|}{T}.  \label{Nbulk}
\end{equation}
Through this paper we set $k_{B}=1=c=\hbar $ for simplicity. We also assume
the temperature associated with the Hubble horizon is the Hawking
temperature $T=H/2\pi $, and the energy contained inside the Hubble volume $%
V=4\pi /3H^{3}$ is the Komar energy
\begin{equation}
E_{\mathrm{Komar}}=|(\rho +3p)|V.  \label{Komar}
\end{equation}
The novel idea of Padmanabhan is that the cosmic expansion, conceptually
equivalent to the emergence of space, is being driven towards holographic
equipartition, and the basic law governing the emergence of space must
relate the emergence of space to the difference between the number of
degrees of freedom in the holographic surface and the one in the emerged
bulk \cite{Pad1}. He proposed that in an infinitesimal interval $dt$ of
cosmic time, the increase $dV$ of the cosmic volume is given by
\begin{equation}
\frac{dV}{dt}=L_{p}^{2}(N_{\mathrm{sur}}-N_{\mathrm{bulk}}).  \label{dV}
\end{equation}
In general, one may expect ${dV}/{dt}$ to be some function of $(N_{\mathrm{%
sur}}-N_{\mathrm{bulk}})$ which vanishes when the latter does. In this case
one may regard Eq. (\ref{dV}) as a Taylor series expansion of this function
truncated at the first order \cite{Pad1}. This issue was also investigated
in \cite{Yang}. Substituting the cosmic volume $V=4\pi /3H^{3}$, the degrees
of freedom on the holographic boundary (\ref{Nsur}), the temperature $%
T=H/2\pi $, and the bulk degrees of freedom in Eq. (\ref{dV}),
after using the Komar energy in the bulk (\ref{Komar}), the
standard dynamical equation for the Friedmann model can be
obtained \cite{Pad1}
\begin{equation}
\frac{\ddot{a}}{a}=-\frac{4\pi L_{p}^{2}}{3}(\rho +3p).  \label{ddota}
\end{equation}
Multiplying both hand side of (\ref{ddota}) by factor $\dot{a}a$ and using
the continuity equation
\begin{equation}
\dot{\rho}+3H(\rho +p)=0,  \label{cont}
\end{equation}
after integrating, one obtains the Friedmann equation
\begin{equation}
H^{2}+\frac{k}{a^{2}}=\frac{8\pi L_{p}^{2}}{3}\rho ,  \label{Fr1}
\end{equation}
where $k$ is an integration constant, which can be interpreted as the
spatial curvature of the FRW universe.

It is important to note that in the above derivation, the number of degrees
of freedom on the holographic boundary (\ref{Nsur}) as well as the
entropy-area relation play a crucial role. However, the entropy-area
relation can be modified from the inclusion of quantum effects \cite
{Shey1,Shey2}. To consider the general corrections to area law, we write
down the general entropy-corrected relation as \cite{Hendi1}
\begin{equation}
S=\frac{A}{4L_{p}^{2}}+{s}(A),  \label{S}
\end{equation}
where $s(A)$ stands for the general correction terms. Applying the general
entropy formula (\ref{S}) to the holographic surface, we assume that the
effective area of the holographic surface is
\begin{equation}
\widetilde{A}=A+4L_{P}^{2}s(A),
\end{equation}
where $A=4\pi H^{-2}$. Using the expression for the volume $V=4\pi /3H^{3}$
as well as the area $A$ of a $3$-sphere with radius $R$, one has
\begin{equation}
\frac{dV}{dA}=\frac{1}{2H},  \label{dVdA}
\end{equation}
and
\begin{equation}
\frac{dA}{dt}=-8\pi H^{-3}\dot{H}.  \label{dAt}
\end{equation}
We also assume that the effective volume enveloped by the
effective holographic surface $\widetilde{A}$ increase with time
according to
\begin{eqnarray}  \label{dVtil}
\frac{d\widetilde{V}}{dt}=\frac{d\widetilde{V}}{d\widetilde{A}}\frac{d%
\widetilde{A}}{dt} &=&\frac{1}{2H}\left( \frac{dA}{dt}+4L_{p}^{2}\frac{ds(A)%
}{dA}\frac{dA}{dt}\right) ,  \label{dVtil1} \\
&=&-4\pi H^{-4}\dot{H}\left( 1+4L_{p}^{2}\frac{ds(A)}{dA}\right) , \\
&=&-2\pi H^{-5}\left(
2\dot{H}H+8L_{p}^{2}\frac{ds(A)}{dA}H\dot{H}\right).
\end{eqnarray}
The above equation can be further rewritten
\begin{equation}
\frac{d\widetilde{V}}{dt}=-2\pi H^{-5}\frac{d}{dt}\left(
H^{2}+8L_{p}^{2}\int \frac{ds(A)}{dA}Hd{H}\right) .  \label{dVtil2}
\end{equation}
Inspired by Eq. (\ref{dVtil2}), we propose that the number of
degrees of freedom on the holographic surface with general entropy
expression (\ref{S}) is given by
\begin{equation}
N_{\mathrm{sur}}=\frac{4\pi }{L_{p}^{2}H^{4}}\left(
H^{2}+8L_{p}^{2}\int \frac{ds(A)}{dA}Hd{H}\right).  \label{Nsur2}
\end{equation}
The bulk degrees of freedom is still given by (\ref{Nbulk}),
\begin{eqnarray}
N_{\mathrm{bulk}} &=&\frac{-4\pi (\rho +3p)V}{H},  \notag  \label{Nbulk2} \\
&=&-\frac{16\pi ^{2}}{3}H^{-4}(\rho +3p),
\end{eqnarray}
where we have added a minus sign in front of $E_{\mathrm{Komar}}$, in order
to have $N_{\mathrm{bulk}}>0$, which makes sense only in an accelerating
universe with $\rho +3p<0$ \cite{Pad1}. Next, motivated by Eq. (\ref{dV}),
we assume
\begin{equation}
\frac{d\widetilde{V}}{dt}=L_{p}^{2}(N_{\mathrm{sur}}-N_{\mathrm{bulk}}).
\label{dVtil3}
\end{equation}
Substituting Eqs. (13), (\ref{Nsur2}) and (\ref{Nbulk2}) in relation (\ref
{dVtil3}), after simplification we get
\begin{equation}
\frac{\ddot{a}}{a}+4L_{p}^{2}\left( \frac{ds(A)}{dA}\dot{H}+2\int \frac{ds(A)%
}{dA}Hd{H}\right) =-\frac{4\pi L_{p}^{2}}{3}(\rho +3p).  \label{dda2}
\end{equation}
This is nothing, but the general form of the entropy corrected dynamical
equation of a FRW universe filled by perfect fluid. To simplify the above
equation further we need to identify the functional form of the correction
term $s(A)$. This is the subject of the next section. In the absence of the
correction term, $s(A)=0$, and Eq. (\ref{dda2}) reduces to the well-known
standard dynamical equation (\ref{ddota}) of FRW universe.
%%%%%%%%%%%%%%%%%%%%%%%%%%%%%%%%%%%%%%%%%%%%%%%%%%%%%%%%%%%%%%%%%%
\section{Emergence of Entropy corrected Friedmann equations\label{entr}}
In this section we apply the general method developed in the
previous section to derive the dynamical Friedmann equations in
the presence of the correction terms to the area law of entropy.
Two well-known quantum corrections to the area law have been
introduced in the literatures, namely, logarithmic and power-law
corrections. Logarithmic corrections, arises from the loop quantum
gravity due to thermal equilibrium fluctuations and quantum
fluctuations \cite{Log}
\begin{equation}
S=\frac{A}{4L_{p}^{2}}+\alpha \ln \frac{A}{4L_{p}^{2}}+\beta \frac{4L_{p}^{2}%
}{A},  \label{Slog}
\end{equation}
where $\alpha $ and $\beta $ are dimensionless constants of order
unity. The exact values of these constants are not yet determined
and still is an open issue in loop quantum cosmology (see
\cite{Cai2} and references therein). These corrections arise in
the black hole entropy in loop quantum gravity due to thermal
equilibrium fluctuations and quantum fluctuations
\cite{Rovelli,Zhang}. The logarithmic term also appears in a model
of entropic cosmology which unifies the inflation and late time
acceleration \cite{Cai3}. Another form of correction to area law,
namely the power-law correction, appears in dealing with the
entanglement of quantum fields inside and outside the horizon. The
entanglement entropy of the ground state obeys the well-known area
law. Only the excited state contributes to the correction, and
more excitations produce more deviation from the area law
\cite{sau1}. For a recent review on the origin of black hole
entropy through entanglement, see \cite{sau2}. The power-law
corrected entropy can be written as \cite{Sau,pavon1}
\begin{equation}
S=\frac{A}{4L_{p}^{2}}\left[ 1-K_{\alpha }A^{1-\alpha /2}\right] ,
\label{Spl}
\end{equation}
where $\alpha $ is a dimensionless constant whose value is currently under
debate, and
\begin{equation}
K_{\alpha }=\frac{\alpha (4\pi )^{\alpha /2-1}}{(4-\alpha )r_{c}^{2-\alpha }}%
,  \label{Kalpha}
\end{equation}
where $r_{c}$ is the crossover scale. The second term in Eq.
(\ref{Spl}) can be regarded as a power-law correction to the area
law, resulting from entanglement, when the wave-function of the
field is chosen to be a superposition of ground state and exited
state \cite{Sau}.

Let us first consider the logarithmic corrected entropy (\ref{Slog}). We
define the effective area of the holographic surface corresponding to
entropy (\ref{Slog}) as
\begin{eqnarray}
\widetilde{A} &=&4L_{p}^{2}\ S \\
&=&4L_{p}^{2}\left[ \frac{A}{4L_{p}^{2}}+\alpha \ \ln \frac{A}{4L_{p}^{2}}%
+\beta \frac{4L_{p}^{2}}{A}\right] .
\end{eqnarray}
Now we calculate the increasing in the effective volume,
\begin{equation*}
\frac{d\widetilde{V}}{dt}=\frac{1}{2H}\frac{d\widetilde{A}}{dt}=\frac{1}{2H}%
\left[ 1+\alpha \frac{4L_{p}^{2}}{A}-\beta \frac{16L_{p}^{4}}{A^{2}}\right]
\frac{dA}{dt}.
\end{equation*}
Using relation $A=4\pi H^{-2}$, after some simplification we get
\begin{eqnarray}
\frac{d\widetilde{V}}{dt} &=&4\pi \dot{H}H^{-4}\left( -1-\frac{\alpha
L_{p}^{2}}{\pi }H^{2}+\frac{\beta L_{p}^{4}}{\pi ^{2}}H^{4}\right) ,
\label{dVtillog1} \\
&=&-2\pi H^{-5}\frac{d}{dt}\left( H^{2}+\frac{\alpha L_{p}^{2}}{2\pi }H^{4}-%
\frac{\beta \ L_{p}^{4}}{3\pi ^{2}}H^{6}\right) .  \label{dVtillog2}
\end{eqnarray}
Next we suppose from (\ref{dVtillog2}) that the number of degrees of freedom
on the holographic surface is given by
\begin{equation}
N_{\mathrm{sur}}=\frac{4\pi }{L_{p}^{2}H^{4}}\left[ H^{2}+\frac{\alpha
L_{p}^{2}}{2\pi }H^{4}-\frac{\beta L_{p}^{4}}{3\pi ^{2}}H^{6}\right] .
\label{Nsurlog}
\end{equation}
Substituting Eqs. (\ref{Nbulk2}), (\ref{dVtillog1}) and (\ref{Nsurlog}) in
relation (\ref{dVtil3}), one obtains
\begin{equation}
\dot{H}\left( 1+\frac{\alpha L_{p}^{2}}{\pi }H^{2}-\beta \frac{L_{p}^{4}}{%
\pi ^{2}}H^{4}\right) +H^{2}+\frac{\alpha L_{p}^{2}}{2\pi }H^{4}-\beta \frac{%
L_{p}^{4}}{3\pi ^{2}}H^{6}=-\frac{4\pi L_{p}^{2}}{3}(\rho +3p).
\end{equation}
Using the continuity equation (\ref{cont}) and multiplying the both hand
side by factor $2\dot{a}a$, the above equation can be written
\begin{equation}
\frac{d}{dt}\left( H^{2}a^{2}\right) +\frac{\alpha L_{p}^{2}}{2\pi }\frac{d}{%
dt}(H^{4}a^{2})-\beta \frac{L_{p}^{4}}{3\pi ^{2}}\frac{d}{dt}\left(
H^{6}a^{2}\right) =\frac{8\pi L_{p}^{2}}{3}\frac{d}{dt}\left( \rho
a^{2}\right) .  \label{Frlog}
\end{equation}
Integrating (\ref{Frlog}), we arrive at
\begin{equation}
H^{2}+\frac{\alpha L_{p}^{2}}{2\pi }H^{4}-\frac{\beta L_{p}^{4}}{3\pi ^{2}}%
H^{6}=\frac{8\pi L_{p}^{2}}{3}\rho ,  \label{Frlog2}
\end{equation}
where we have put the integration constant equal to zero. This is
exactly the result of \cite{Cai2,Sheykhi1} for the entropy
corrected Friedmann equation derived by applying the first law of
thermodynamics on the apparent horizon of FRW universe. Here we
arrived at the same result by using completely different approach.
This indicates that, given the entropy expression at hand, one is
able to reproduce the corresponding dynamical equation using the
novel Padmanabhan's idea on the emergence nature of spacetime
dynamics due to the difference between the surface degrees of
freedom and the bulk degrees of freedom in a region of space. Note
that, unlike the case of Einstein's gravity, one cannot interpret
the constant of integration as the spatial curvature of the
universe if one does not set it equal to zero \cite{Cai1}.

Before we proceed further, let us compare the result obtained here with the
modified Friedmann equation in loop quantum cosmology. Applying the
techniques of loop quantum gravity to homologous and isotropic spacetime
leads to the so-called loop quantum cosmology. Due to quantum correction,
the Friedmann equations get modified. The big bang singularity is resolved
and replaced by a quantum bounce \cite{Ash1}. For a brief summary on loop
quantum cosmology, see \cite{Ash2}. Considering quantum correction, the
modified Friedmann equation for spatially flat universe turns out to be
\begin{equation}  \label{Frloop}
H^2=\frac {8\pi L_{p}^2}{3} \rho \left(1-\frac{\rho}{\rho _{\mathrm{crit}}}%
\right),
\end{equation}
where
\begin{equation}  \label{rhocri1}
\rho _{\mathrm{crit}}=\frac{\sqrt{3}}{(32 \pi L_{p}^4 \gamma^3)},
\end{equation}
is the critical energy density. Here $\gamma$ is Barbero-Immirzi parameter
which could be fixed as $0.2375$ in order to give the area formula of black
hole entropy in loop quantum gravity \cite{Ash3}. Due to the corrected term
in (\ref{Frloop}), the big bang singularity is replaced by a quantum bounce
happening at $\rho=\rho _{\mathrm{crit}}$.

Let us stress here that although in the literature there is doubt about the
second correction term in entropy-corrected relation (\ref{Slog}), it is
widely believed that the first correction term originates from loop quantum
gravity is the logarithmic term. Considering the case with $\beta =0$, the
modified Friedmann equation (\ref{Frlog2}) reduces to
\begin{equation}
H^{2}+\frac{\alpha L_{p}^{2}}{2\pi }H^{4}=\frac{8\pi L_{p}^{2}}{3}\rho .
\label{Frloop2}
\end{equation}
Solving for $H^{2}$ we have
\begin{equation}
H^{2}=\frac{\pi }{\alpha L_{p}^{2}}\left( -1+\sqrt{1+\frac{16\alpha L_{p}^{4}%
}{3}\rho }\right) .  \label{Frloop3}
\end{equation}
If we regard $\alpha $ as a small quantity, we can expand the right hand
side of (\ref{Frloop3}), up to the linear order of $\alpha $ and get
\begin{equation}
H^{2}\approx \frac{8\pi L_{p}^{2}}{3}\rho \left( 1-\frac{4\alpha L_{p}^{4}}{3%
}\rho \right) .  \label{Frloop4}
\end{equation}
It is apparent that Eq. (\ref{Frloop4}) is quite similar to the modified
Friedmann equation (\ref{Frloop}) in loop quantum cosmology, provided we
define
\begin{equation}
\rho _{\mathrm{crit}}=\frac{3}{4\alpha L_{p}^{4}},  \label{rhocri2}
\end{equation}
which is in agreement with (\ref{rhocri1}) if we define $\alpha =8\sqrt{3}%
\pi \gamma ^{3}$. For $\gamma =0.2375$, one gets $\alpha \simeq 0.58$.

Next, we consider the power-low correction to the entropy relation, namely (%
\ref{Spl}). In this case, by using the same procedure, it is easy
to show that the effective volume varies with time according to
\begin{eqnarray}  \label{dVpl}
\frac{d\widetilde{V}}{dt}=-2 \pi H^{-5} \frac{d}{dt}\left(H^{2}-\frac{%
H^{\alpha}}{r_c^{2-\alpha}}\right),
\end{eqnarray}
where we have also used relation (\ref{Kalpha}). Therefore,
inspired by (\ref{dVpl}), we assume the number of degrees of
freedom on the holographic surface with radius $H^{-1}$ is given
by
\begin{equation}  \label{Nsurpl}
N_{\mathrm{sur}}=\frac{4\pi}{L_{p}^{2}H^{4}}\left(H^{2}-\frac{H^{\alpha}}{%
r_c^{2-\alpha}}\right).
\end{equation}
The number of degrees of freedom inside the bulk is still given by relation (%
\ref{Nbulk2}). Combining Eqs. (\ref{Nbulk2}), (\ref{dVpl}) and (\ref{Nsurpl}%
) with relation (\ref{dVtil3}), after using continuity equation
(\ref {cont}), we arrive at
\begin{equation}
\frac{d}{dt}\left(H^{2}a^{2}\right)- r^{\alpha-2}_{c} \frac{d}{dt}%
\left(H^{\alpha} a^{2}\right)=\frac{8\pi L_{p}^2}{3}\frac{d}{dt}\left(\rho
a^{2}\right).
\end{equation}
Integrating and dividing by $a^{2}$ we reach
\begin{equation}  \label{Frpl}
H^{2}-r^{\alpha-2}_{c}H^{\alpha}=\frac{8\pi L_{p}^{2}}{3}\rho,
\end{equation}
where we have again set the integration constant equal to zero. This
equation is indeed the corresponding Friedmann equation of a flat FRW
universe in the presence of power-law corrected entropy derived in \cite
{Shey2}. Note that in \cite{Shey2} one of the present author and Hendi
derived Eq. (\ref{Frpl}) by applying the first law of thermodynamics on the
apparent horizon (see Eq. (47) of \cite{Shey2}), however our approach here
is quite different. Here we arrived at (\ref{Frpl}) by assuming that the
difference between the number of degrees of freedom in the bulk and on the
boundary is proportional to the volume change of the spacetime. This is a
remarkable result and shows that the approach presented here is powerful
enough to derive the correct form of the gravitational field equations.
%%%%%%%%%%%%%%%%%%%%%%%%%%%%%%%%%%%%%%%%%%%%%%%%%%%%%%%%%%%%%%%%%%%%%%%%%

\section{Emergence of Friedmann equations in braneworld\label{brane}}

In the previous section we showed that using the Padmanabhan's new idea one
is able to re-produce the dynamical equations of FRW universe by calculating
the number of degrees of freedom on the Hubble horizon and inside the
universe. Is the obtained relation between the degrees of freedom and the
gravitational field equation just an accident? Does this connection hold in
all dynamical spacetimes and even beyond Einstein gravity? Does it imply
something in deep? Very recently, Cai showed that the above procedure can be
applied to extract the corresponding Friedmann equation of flat FRW universe
in Gauss-Bonnet and more general Lovelock gravity\cite{Cai1}. In this
section we will address the question on the connection between the degrees
of freedom and spacetime dynamics by investigating whether and how the
relation can be found in braneworld models. Gravity on the brane does not
obey Einstein theory, thus the usual area formula for the holographic
boundary get modified on the brane \cite{SheyW1,SheyW2}.

There are two main pictures in the braneworld scenario. In the first picture
which we refer as the RS II model, a positive tension 3-brane embedded in an
5-dimensional AdS bulk and the cross over between 4D and 5D gravity is set
by the AdS radius \cite{RS,Bin}. In this case, the extra dimension has a
finite size and the localization of gravity on the brane occurs due to the
negative cosmological constant in the bulk. In another picture which is
based on the work of DGP model \cite{DGP,DG}, a 3-brane is embedded in a
spacetime with an infinite-size extra dimension, with the hope that this
picture could shed new light on the standing problem of the cosmological
constant as well as on supersymmetry breaking \cite{DGP,Wit}. The recovery
of the usual gravitational laws in this picture is obtained by adding to the
action of the brane an Einstein-Hilbert term computed with the brane
intrinsic curvature.

Let us begin by RS II model in which no intrinsic curvature term on the
brane contributes in the action. The entropy associated with the apparent
horizon for an $(n-1)$-brane embedded in an $(n+1)$-dimensional bulk in RS
II brane scenario is given by \cite{SheyW1}
\begin{equation}  \label{entRSAdS1}
S=\frac{2\Omega_{n-1}{\tilde{r}_A}^{n-1}}{4 G_{n+1}} \times {}_2F_1\left(%
\frac{n-1}{2},\frac{1}{2},\frac{n+1}{2}, -\frac{{\tilde{r}_A}^2}{\ell^2}%
\right),
\end{equation}
where ${}_2F_1(a,b,c,z)$ is a hypergeometric function, and $\ell$ is the
bulk AdS radius,
\begin{equation}  \label{rela}
\ell^2=-\frac{n(n-1)}{16\pi G_{n+1} \Lambda_{n+1}}\, ,\quad \Omega_{n-1}=%
\frac{\pi^{(n-1)/2}}{\Gamma((n+1)/2)}.
\end{equation}
Here $\Lambda_{n+1}$ is the $(n+1)$-dimensional bulk cosmological constant
and $\tilde{r}_A$ is the apparent horizon radius which can be extended into
the bulk. The explicit evaluation of the apparent horizon for the FRW
universe gives \cite{SheyW1}
\begin{equation}  \label{radius}
\tilde{r}_A=\frac{1}{\sqrt{H^2+k/a^2}}.
\end{equation}
It is worth noticing when $\tilde{r}_A \ll\ell$, which physically
means that the size of the extra dimension is very large if
compared with the apparent horizon radius, one recovers the
$(n+1)$-dimensional area formula for the entropy on the brane, $S
=2\Omega_{n-1}{\tilde{r}_A}^{n-1}/4G_{n+1}$. The factor $2$ comes
from the $\mathbb{Z}_2$ symmetry in the bulk \cite{SheyW1}. This
is due to the fact that in this case $\Lambda_{n+1}\rightarrow 0$
and thus in the absence of the negative cosmological constant in
the bulk, no localization of gravity happens on the brane. As a
result, the gravity on the brane is still $(n+1)$-dimensional.

We further rewrite the entropy expression (\ref{entRSAdS1}) in the form \cite
{SheyW1}
\begin{equation}  \label{entRSAdS2}
S=\frac{(n-1)\ell \Omega_{n-1}}{2G_{n+1}}{\displaystyle\int^{\tilde r_A}_0%
\frac{\tilde{r}_A^{n-2} }{\sqrt{\tilde{r}_A^2+\ell^2}}d\tilde{r}_A}.
\end{equation}
Now we consider the flat $4$-dimensional universe (the generalization of
this study to arbitrary dimensions is quite straightforward). In this case
we have $n=4$ and $\tilde{r}_A=H^{-1}$ and relation (\ref{entRSAdS2}) can be
written as
\begin{equation}
S=\frac{3 \ell \Omega_{3}}{2 G_{5}}\int\frac{H^{-2}}{\sqrt{H^{-2}+\ell^{2}}}%
d(H^{-1}).
\end{equation}
We define the effective area as
\begin{equation}
\widetilde{A}=4G_{5} S=6 \ell\Omega_{3}\int\frac{H^{-2}}{\sqrt{%
H^{-2}+\ell^{2}}}d(H^{-1}).
\end{equation}
Taking the time derivative, we get
\begin{eqnarray}
\frac{d\widetilde{A}}{dt}&=&6 \ell\Omega_{3}\frac{d}{dt} \left(\int\frac{%
H^{-2}}{\sqrt{H^{-2}+\ell^{2}}}d(H^{-1})\right) \\
&=&6 \ell\Omega_{3}\frac{d}{d(H^{-1})} \left(\int\frac{H^{-2}}{\sqrt{%
H^{-2}+\ell^{2}}}d(H^{-1})\right)\frac{d(H^{-1})}{dt}.
\end{eqnarray}
Using the fact that ${d(H^{-1})}/{dt}=-\dot{H}H^{-2}$, we arrive at
\begin{equation}
\frac{d\widetilde{A}}{dt}=-6 \ell\Omega_{3}\frac{H^{-4}\dot{H}}{\sqrt{%
H^{-2}+\ell^{2}}}.
\end{equation}
Therefore, the increase in the effective volume is obtained as
\begin{eqnarray}  \label{dVRS0}
\frac{d\widetilde{V}}{dt}&=&-3 \Omega_{3}\frac{H^{-4}\dot{H}}{\sqrt{%
H^{2}+1/\ell^{2}}}.
\end{eqnarray}
Next, we assume the number of degrees of freedom on the apparent horizon is
given by
\begin{equation}  \label{NsurRS}
N_{\mathrm{sur}}=\frac{6\Omega_{3}}{G_{5}}H^{-4}\sqrt{H^{2}+\frac{1}{\ell^{2}%
}}.
\end{equation}
We also replace $L_{p}^2$ in relation (\ref{dVtil3}), with $G_{5}$,
\begin{equation}  \label{dVRS}
\frac{d\widetilde{V}}{dt}=G_{5}(N_{\mathrm{sur}} - N_{\mathrm{bulk}}).
\end{equation}
Inserting Eqs. (\ref{Nbulk2}), (\ref{dVRS0}) and (\ref{NsurRS}) in relation (%
\ref{dVRS}), one gets
\begin{equation}
\frac{\dot{H}}{\sqrt{H^2+1/\ell^{2}}}+2 \sqrt{H^{2}+\frac{1}{\ell^{2}}}=-%
\frac{4\pi G_{5}}{3}(\rho+3p).
\end{equation}
Using the continuity equation (\ref{cont}) and multiplying the both hand
side by $\dot{a}a$, we arrive at
\begin{equation}
\frac{d}{dt}\left(a^{2}\sqrt{H^{2}+\frac{1}{\ell^{2}}}\right)=\frac{4\pi
G_{5}}{3}\frac{d}{dt}\left(\rho a^{2}\right).
\end{equation}
Integrating and dividing by $a^{2}$ we obtain
\begin{equation}
\sqrt{H^{2}+\frac{1}{\ell^{2}}}=\frac{4\pi G_{5}}{3}\rho,
\end{equation}
where we set the integration constant equal to zero. In this way we derive
the Friedmann equation of flat FRW universe in RS II by determining the
difference between the number of degrees of freedom on the boundary and in
the bulk (see Eq. (34) in \cite{SheyW1}). The main characteristic of
Friedmann equation in RS II braneworld model, namely $H^2\propto \rho^2$ can
be seen in obtained equation. If one invokes the standard assumption that
the energy density on the brane can be separated into two contributions, the
ordinary matter component, $\rho_{b}$, and the brane tension, $\lambda > 0$,
such that $\rho= \rho_{b}+\lambda$, (after fine tuning between the brane
tension and the bulk cosmological constant), then one can recover another
form of the Friedmann equation in RS II braneworld \cite{CaiCao}.

Finally, we extend the discussion to the case in which the
intrinsic curvature term of the brane is included in the action,
namely DGP braneworld. In this case, the entropy of the apparent
horizon for an $(n-1)$-brane embedded in an $(n+1)$-dimensional
Minkowski bulk is given by \cite{SheyW1}
\begin{eqnarray}  \label{entDGmin1}
S &=&\frac{(n-1)\Omega_{n-1}{\tilde{r}_A}^{n-2}}{4G_n}-\epsilon\frac{%
2\Omega_{n-1}{\tilde{r}_A}^{n-1}} {4G_{n+1}}=S_{n}-\epsilon S_{n+1}.
\end{eqnarray}
where $\epsilon =\pm 1$ corresponds to the two branch of the DGP
braneworld \cite{Def}. It is interesting to note that in this case
the entropy can be regarded as a sum of two area formulas; one
(the first term) corresponds to the $4$-dimensional gravity on the
brane and the other (the second term) to the $5$-dimensional
gravity in the bulk. The factor $2$ in the second term comes from
the $\mathbb{Z}_2$ symmetry in the bulk \cite{DGP}. This indeed
reflects the fact that there are two gravity terms in the action
of DGP model. Consider
flat $4$-dimensional FRW universe on the brane, with $n=4$ and $\tilde{r}%
_A=H^{-1}$, we have
\begin{equation}
S=\frac{3\Omega_{3}}{4G_{4}}H^{-2}-\epsilon\frac{2\Omega_{3}}{4G_{5}}H^{-3}.
\end{equation}
Defining the effective surface, we have
\begin{equation}
\widetilde{A}=4G_{5}S=2\Omega_{3}\left(\frac{3G_{5}}{2 G_{4}}H^{-2}-\epsilon
H^{-3}\right),
\end{equation}
which varies with time according to
\begin{equation}
\frac{d\widetilde{A}}{dt}=-6\Omega_{3}\dot{H}H^{-3}\left(\frac{G_{5}}{G_{4}}%
-\epsilon H^{-1}\right).
\end{equation}
Hence, the effective volume increase is obtained as
\begin{eqnarray}  \label{dVDGP1}
\frac{d\widetilde{V}}{dt}=\frac{1}{2H}\frac{d\widetilde{A}}{dt}&=&-3\dot{H}%
H^{-4}\Omega_{3}\left(\frac{G_{5}}{G_{4}}-\epsilon H^{-1}\right) \\
&=&3\Omega_{3}\frac{d}{dt}\left(\frac{G_{5}}{3G_{4}}H^{-3}-\epsilon\frac{%
H^{-4}}{4}\right).
\end{eqnarray}
In this case we assume the number of degrees of freedom to be
\begin{equation}  \label{NsurDGP}
N_{\mathrm{sur}}=\frac{3\Omega_{3}H}{G_{5}}\left(\frac{G_{5}}{G_{4}}%
H^{-3}-2\epsilon H^{-4}\right).
\end{equation}
Substituting (\ref{dVDGP1}) and (\ref{NsurDGP}) into (\ref{dVRS}) together
with the bulk degrees of freedom (\ref{Nbulk2}), we arrive at
\begin{equation}
\frac{G_{5}}{2G_{4}}\left(2\dot{H}+2H^{2}\right)-\epsilon H^{-1}\left(\dot{H}%
+2H^{2}\right)=-\frac{4 \pi G_5}{3}(\rho+3p).
\end{equation}
Integrating this equation with help of the continuity equation yields
\begin{equation}  \label{FrDGP1}
\frac{G_{5}}{2G_{4}}H^{2}-\epsilon H =\frac{4\pi G_{5}}{3}\rho,
\end{equation}
where we have assumed the integration constant to be zero. We define the
crossover length scale between the small and large distances in DGP
braneworld as \cite{Def}
\begin{equation}
r_c=\frac{G_5}{2G_4}.
\end{equation}
Thus Eq. (\ref{FrDGP1}) can be rewritten
\begin{equation}  \label{FrDGP2}
H^{2}-\frac{\epsilon}{r_{c}}H =\frac{8\pi G_{4}}{3}\rho.
\end{equation}
The above equation can be further rewritten in the following form
\begin{equation}  \label{FrDGP3}
H^2=\Bigg[\frac{\epsilon}{2r_c}+\sqrt{\frac{8\pi G_{4}}{3}\rho+\frac{1}{%
4r_c^2}}\,\,\Bigg]^2.
\end{equation}
This equation is indeed the corresponding Friedmann equation of flat FRW
universe in DGP brane scenario derived in \cite{Def} from the field
equations. When $r_c\rightarrow \infty$, one recovers the standard Friedmann
equation in flat FRW universe. Physically, this means that the apparent
horizon is not extended in the bulk and located totally on the brane. As a
result, the effect of the extra dimension, does not appear in the Friedmann
equation. Equation (\ref{FrDGP2}) with $\epsilon=1$ and $\rho=0$ has an
interesting self-accelerating solution with a Hubble parameter given by the
inverse of the crossover scale $r_c$ \cite{Def1}. This is due to the fact
that the intrinsic curvature term on the DGP brane appears as a source for
the bulk gravity, so that with appropriate initial conditions this term can
cause an expansion of the braneworld without the need of matter or a
cosmological constant on the brane \cite{Def1}.

\section{Closing remarks\label{con}}

In summary, we have investigated the novel idea recently proposed by
Padmanabhan \cite{Pad1} which argues that the acceleration of the cosmic
expansion is due to the difference between the number of degrees of freedom
on the Hubble horizon of the universe and the one in the emerged bulk. This
new proposal also leads to derive the dynamical equation governing the
evolution of the universe. Since in general, the entropy associated with the
horizon is a function of its area and depends on the underlying theory of
gravity, therefore any modification of the entropy expression leads to a
particular number of degrees of freedom on the Hubble horizon.

In this paper, we developed the method of \cite{Pad1} by assuming
that the entropy associated with the horizon is a general function
of its area. In this case, we derived successfully the general
expression for the number of degrees of freedom on the Hubble
horizon as well as the general expression for the effective volume
increase. We considered two special corrections to the entropy
expression namely, power-law and logaritmic corrections and
extracted the corresponding modified Friedmann equations in flat
FRW universe. We also applied our general formalism to the
braneworld scenarios. Using the entropy expression associated with
the apparent horizon on the brane, we calculated the volume
increase and the number of degrees of freedom on the apparent
horizon on the brane. Then, we obtained the Friedmann equations in
RS II and DGP braneworld models. Our study may indicate that the
novel proposal of Padmanabhan \cite{Pad1} is powerful enough to
apply for deriving the dynamical equations in other gravity
theories.

It is important to note that in this work, we only modified the
number of degrees of freedom on the Hubble horizon, $N_{\rm sur}$,
while we keep the number of degrees of freedom in the bulk,
$N_{\rm bulk}$, as in case of standard cosmology. Indeed, when the
underlying theory of gravity is modified, or the entropy
expression get quantum correction terms, then $N_{\rm sur}$ get
modified as well. However, we proposed the bulk degrees of
freedom, $N_{\rm bulk}$, has the same expression as in the absence
of correction terms. This is due to the fact that we have assumed
$N_{\rm bulk}$  depends only on the matter degrees of freedom.

Finally, we would like to mention that unlike the case of Einstein
gravity, we cannot interpret the integration constant as the
spatial curvature of the FRW universe. In other words, we could
only derive the Friedmann equations in flat universe. Thus, it is
of great importance to extend the method developed in this paper
for the spatially non flat FRW universe and derive the
corresponding Friedmann equations. This is quite an interesting
subject, which deserves further investigation.

%%%%%%%%%%%%%%%%%%%%%%%%%%%%%%%%%%%%%%%%%%%%%%%%%%%%%%%%%%%%%%%%%%%%%%%
\acknowledgments{We thank the referee for constructive comments
which helped us to improve the paper. We also acknowledge from the
Research Council of Shiraz University. This work has been
supported financially by Center for Excellence in Astronomy and
Astrophysics (CEAA-RIAAM), Maragha.}
%%%%%%%%%%%%%%%%%%%%%%%%%%%%%%%%%%%%%%%%%%%%%%%%%%%%%%%%%%%%%%%%%%%%%%%

\end{document}